\documentclass[preprint,prd]{revtex4}
\usepackage{amstext,amsmath,amssymb,amsfonts,bm,bbm}
\usepackage{euscript}
\usepackage[usenames,dvipsnames]{color}
\usepackage{graphicx}

\def\be{\begin{equation}}
\def\ee{\end{equation}}
\def\bal{\begin{align}}
\def\eal{\end{align}}
\def\bay{\begin{array}}
\def\ear{\end{array}}

\def\0{\otimes}
\def\1{{{\mathbb 1}}}
\def\6{\langle}
\def\9{\rangle}

\def\hp{\hat {p}}
\def\hq{\hat{q}}
\def\hx{\hat {x}}
\def\hr{\hat{r}}

\def\p{\prime}

\def\ap{\hat{a}^\prime_i}
\def\adp{\hat{a}^{\dagger \p}_i}
\def\bn{\bar{n}_i^\p}
\def\n{n_i^\p}
\def\om{\omega_i^\prime}

\def\e0{\epsilon_0}

\begin{document}

\title{Pedagogical introduction to the entropy of entanglement\\for Gaussian states.}
\author{Tommaso F. Demarie}
\affiliation{Department of Physics \& Astronomy, Faculty of Science, Macquarie University, NSW 2109, Australia}

\begin{abstract}
The most useful measure of a bipartite entanglement is  the von Neumann entropy of either of the reduced density matrices. For a particular class of continuous-variable states, the Gaussian states, the entropy of entanglement can be expressed rather elegantly in terms of the \textit{symplectic eigenvalues}, elements that characterize a Gaussian state and depend on the correlations of the canonical variables. We give a pedagogical step-by-step derivation of this result and provide some insights that  can be useful in practical calculations.

\end{abstract}

\maketitle

\section{Introduction}
Entanglement is recognized as a key resource in quantum computation \cite{horos}. A number of protocols 
utilize entanglement for performing tasks that would be very hard, if not impossible, in classical information processing \cite{tasks}. Different quantitative measures 
capture different aspects of entanglement \cite{plenio06}. For pure states there is a preferred way to quantify the entanglement between a subsystem $A$ and its complement $B$. This is the \textit{degree of entanglement} or \textit{entropy of entanglement} \cite{bennett96}, defined as the von Neumann entropy of either of the reduced density operators $\rho_A$ or $\rho_B$ of the state $\rho=\rho_{AB}$. Explicitly, $S =  - \text{tr} \left( \rho_A \log{\rho_A} \right) =  - \text{tr} \left( \rho_B \log{\rho_B} \right)$. While studies of entanglement originally focused on finite-dimensional quantum systems, continuous-variable systems are becoming increasingly important both practically and theoretically \cite{eisert03}. For instance, a special class of continuous variable states, the Gaussian states, 
plays an important role in quantum optics and quantum information processing and can be described conveniently by an easy algebraic formalism. Furthermore, entanglement calculations in infinite-dimensional systems are much more demanding than their finite-dimensions counterparts. Unlike the general case, the von Neumann entropy of Gaussian states has a simple expression in terms of a finite number of the symplectic eigenvalues $\{\sigma_i\}$ of the correlation matrix of  the quadrature operators. Specifically \cite{entop},
\begin{align}
\label{entropy}
S(\rho)  = \sum_{i=1}^{n_{sub}} \left[ \left(\sigma_i + \frac12 \right) \log_2\left({\sigma_i + \frac12}\right) - \left(\sigma_i - \frac12 \right) \log_2\left({\sigma_i - \frac12}\right) \right] \, ,
\end{align}
where the index $i$ runs over the modes of one of the two subsystems under examination. In this paper we give a simple but precise derivation of this formula, showing the link between covariance matrix of a Gaussian state, symplectic eigenvalues and entropy. First we present a short introduction to Gaussian states, defining a covariance matrix and its symplectic eigenvalues, and then we give a derivation of (\ref{entropy}). We conclude with some observations on entropy and temperature.

\section{Gaussian states}
Gaussian states are the basic ingredient of continuous-variable quantum computation \cite{zhang06, nick06}. They are well-understood, easy to produce in a laboratory and can be described using a simple matrix formalism. Each mode of a quantized electromagnetic field is equivalent to a quantum harmonic oscillator and the quadrature operators $\hq$ and $\hp$ are related to the mode creation and annihilation operators by \cite{cohen}:
\bal
\hat{q} = \frac{1}{\sqrt{2}} \sqrt{\frac{\hbar}{m \omega}} (\hat{a}^\dagger + \hat{a}), \,\,\,\,\,\,\,\,\,\,\,\,\,\,\, \hat{p}=\sqrt{\hbar m \omega} \frac{i}{\sqrt{2}} (\hat{a}^\dagger - \hat{a})\, .
\end{align}
Consider a bosonic system having $2n$ canonical degrees of freedom (such as $n$ light modes), and introduce a $2n$-dimensional column vector $r = (\hq_1, ..., \hq_n, \hp_1, ..., \hp_n)^T$ of quadrature operators. The canonical commutation relations
\be
[ \hq_i, \hp_j ] = i\hbar \delta_{i,j} \,, \, \,\,\,\,\,  [ \hq_i, \hq_j ] = 0 \, ,\, \,\,\,\,\,  [ \hp_i, \hp_j ] = 0 \,,
\ee
can be rewritten in matrix form using the components of $r$ as
\bal
\hbar \, \Omega_{i,j} = - i \left [ r_i, r_j \right] \, , \,\,\,\,\,\,\,\,\,\,\,\,\,\,\,\,\,\,\,\,\,\,\,\, \Omega = \left (
\begin{tabular}{c c}
$0$ & $\mathbb{I}_n$\\
$-\mathbb{I}_n$ & $0$\\
\end{tabular} \right )  \, .
\end{align}
The skew-symmetric matrix $\Omega$ is called symplectic (metric) matrix \cite{simon88,arvind95}. Notice that $\Omega^T = \Omega^{-1} = - \Omega$.\\ 
The ground state of a quantum harmonic oscillator is a Gaussian state. It is clear intuitively from the Gaussian shape of its wave function and will be given a precise meaning shortly. Similarly, the ground state of a system of $n$ harmonic oscillators, whose Hilbert space is given by the tensor product of the individual Hilbert spaces $\mathcal{H} = \bigotimes_{i=1}^n \mathcal{H}_{i}$, is a Gaussian state. In fact, the ground state of any system that is described by a Hamiltonian quadratic in the canonical operators,
\be
\hat{H} = \frac{1}{2} \sum_i^n \sum_j^n r_i H_{i,j} r_j \,,
\ee
defined by a real and positive-semidefinite crossing matrix $H$, is Gaussian \cite{schuch06}. Mathematically this means that the state characteristic function, which we are going to define now, is a Gaussian in the phase space \cite{olivares12}. \\
Any state $\rho$ of $n$ modes can be characterized by the following construction on the phase space of the system. We introduce the Weyl operator
\bal
W_{\eta} = \text{exp}\{- i \eta^T \Omega r \} \equiv \bigotimes_{i=1}^{n} D_i (\alpha_i) = \bigotimes_{i=1}^{n} e^{\alpha_i \hat{a}_i^\dagger - \alpha_i^* \hat{a}_i}\, ,
\end{align}
which in quantum optics is taken as a phase space displacement operator. Here the vector $\eta = (a_1,..., a_n, b_1,..., b_n)^T \in \mathbb{R}^{2 n}$ defines the displacement that can be represented in the complex form as $\alpha_i = \frac{1}{\sqrt{2}} (a_i + i b_i) \in \mathbb{C}$.  The action of the \textit{i}-th mode displacement operator $D_i (\alpha_i)$ on the \textit{i}-th mode ground state results in the coherent state $|\alpha_i\9 = D_i(\alpha_i)|0\9$ \cite{gazeau}.\\
The characteristic function of the state $\rho$ is defined as the expectation value of the Weyl operator (for a textbook reference see for example \cite{puri01}),
\be
\chi_{\rho}(\eta) = \text{tr} [ \rho W_\eta ] \, .
\ee
It is equivalent to the Wigner distribution, which is defined in terms of phase-space variables as
\be
W(q,p) = \frac{1}{\pi^n} \int d^n q^\prime \6 q - q^\prime | \rho | q + q^\prime \9 e^{2 i q^\prime p} \, ,
\ee
and is a phase space representation of the density matrix $\rho$. The Wigner function is usually expressed by the symplectic Fourier transform of the characteristic function,
\be
W(X) = \frac{1}{(2 \pi)^{2n}} \int d^{2 n} \eta e^{i \eta^T \Omega X} \chi_{\rho} (\eta)    \, ,
\ee
with $X = (q_1, ..., q_n, p_1, ..., p_n)^T$. The density operator of the quantum state can be written in terms of its characteristic function by means of a Fourier-Weyl relation
\be
\rho = \frac{1}{(2 \pi)^n} \int d^{2 n} \eta \chi_{\rho} (- \eta) W_{\eta} \, ,
\ee
where $W(\eta)$ is the Weyl operator, and therefore the state is uniquely determined by its characteristic function $\chi_{\rho}$. Finally, a state $\rho$ of $n$ modes is said to be Gaussian whenever the characteristic function has a Gaussian shape in phase space \cite{simon87}, which means that it can be written as
\be
\chi_{\rho} (\eta) = \chi_{\rho} (0) e^{-\frac{1}{4} \eta^T \Omega \Gamma \Omega^T \eta - i D^T \Omega \eta} \, .
\ee
The first two statistical moments are captured by the vector $D = \text{tr} [ \rho \, r_i ] = \6 r_i \9$ of expectation values of the quadrature operators and the $2n \times 2n$ real symmetric matrix $\Gamma$ that carries the information about the variances 
\bal
\Gamma_{i,j} = \text{Re} \, \text{tr} \left[ \rho (r_i -  \6 r_i \9)(r_j -  \6 r_j \9 ) \right] \, .
\end{align}
All higher-order statistical moments of a Gaussian state can be expressed from $D$ and $\Gamma$. The correlation matrix $\Gamma$ is called covariance matrix (or sometimes noise matrix) and it plays a central role in the following discussion of entropy. Local unitary transformations do not change entanglement \cite{plenio06} and hence, since displacements are single modes local translation in phase space, they leave the entanglement properties of the state unaffected \cite{nick11}. Therefore the elements of $D$ contribute nothing to the entanglement and they can all be made zero. Accordingly, we can rewrite the covariance matrix as
\be
\Gamma_{i,j} = \text{Re} \, \text{tr} \left[ \rho \, r_i r_j \right] \, .
\ee
However, the matrix $\Gamma$ cannot be arbitrary. In order for a real symmetric matrix to be the covariance matrix of some physical state we need to fix a constraint: The canonical commutation relations require the positive definiteness of
\be
\Gamma + \frac{1}{2} i \Omega \geqslant 0 \, ,
\ee
which is just another way to rewrite the Heisenberg uncertainty relations \cite{qigs}.\\
Gaussian states are important because of the existence of a class of operators corresponding to common laboratory procedures that preserve the Gaussian properties of the states on which they act. Thus we define a Gaussian unitary operation as a transformation that maps a Gaussian state onto a Gaussian state.\\
There exists a symplectic representation of the Gaussian unitary group \cite{simon87}. To each Gaussian transformation $U$ we can associate a unique symplectic transformation $S \in \text{Sp}(2n, \mathbb{R})$ (for more details about the real symplectic group and its properties see \cite{arvind95}). The group element $S$ describes a linear transformation of the quadrature operators expressed by
\be
\rho^\prime = U (S)\, \rho \, U^{\dagger}(S) \longrightarrow r^\prime = S r = U (S)^{-1} r U(S) \, .
\ee
These transformations preserve the commutation relations. Therefore, the action of any $S\in \text{Sp}(2n, \mathbb{R})$ on the matrix $\Omega$ is given by
\be
\label{om}
i \Omega = [r^\prime_i, r^\prime_j] = S [r_i, r_j] S^T \longrightarrow \Omega = S \Omega S^T \, .
\ee
Furthermore, if $S$ is a symplectic transformation then it also satisfies $S^T = \Omega S^{-1} \Omega^{-1}, S^{-1} = \Omega S^T \Omega^{-1} \in \text{Sp}(2n, \mathbb{R})$. Under the action of a symplectic transformation the covariance matrix transforms as \cite{simon88}
\be
\label{gam}
\Gamma^\prime = \text{cov}\left( S r \right) = S \,\text{cov}(r) \, S^T = S \Gamma S^T \, .
\ee
\subsection{Symplectic eigenvalues}
The next step towards understanding of the entanglement entropy of Gaussian states is to introduce the concept of \textit{symplectic eigenvalues}. In fact, equation (\ref{gam}) tells us that the covariance matrix does not transform by a similarity transformation under the action of the symplectic group. We are therefore interested in finding an alternative form of $\Gamma$ such that its eigenvalues are invariant under a symplectic transformation and can thus uniquely characterize the state. To do so we use Williamson's theorem \cite{will}. It states that any real symmetric positive-definite $2n \times 2n$ matrix, such as the covariance matrix $\Gamma$, can always be made diagonal by means of a suitable symplectic transformation $S_w \in \text{Sp}(2n, \mathbb{R})$,
\be
S_w \Gamma S_w^T = \Gamma_w \, ,
\ee
where now
\be
\Gamma_w = \text{diag}(\sigma_1, \sigma_2, ..., \sigma_n, \sigma_1, \sigma_2, ..., \sigma_n)
\ee
and all the $\sigma_i$ are real. Hence, after the transformation $S_w$ we have, for the transformed canonical operators $\hat{r}^\prime$, that $\text{Re} \6 \hr^\prime_i \, \hr^\prime_j \9= \delta_{i,j} \sigma_i$. The matrix $\Gamma_w$ is called the Williamson normal form of the matrix $\Gamma$. It is important to realize that in general the $\{ \sigma_i \}$ are not the eigenvalues of $\Gamma$ or of any $\Gamma_S = S^T \Gamma S$ determined by a transformation $S$ different from $S_w$.\\
Now define a new matrix $M$ such that $\Gamma = -M \Omega $ and thus
\be
\Gamma \Omega^{-1} = -   M \Omega \Omega^{-1} \to M = \Gamma \Omega \, .
\ee
Using (\ref{om}) it is easy to see that $S^T \Omega = \Omega S^{-1}$ and therefore the action of a symplectic transformation $S$ on the matrix $M$ results in a similarity transformation of $M$ that preserves its eigenvalues,
\be
M^\prime = \Gamma^\prime  \Omega =  S \Gamma S^T  \Omega = S\Gamma \Omega S^{-1} = S M S^{-1} \, .
\ee
Hence every matrix $M$ determined by varying $S$ over the group $\text{Sp}(2n, \mathbb{R})$ shares the same spectrum \cite{simon87}. In particular, if we take the matrix $\Gamma$ to its Williamson form and transform $M$ accordingly, the eigenvalues of the matrix $M^\prime = S_w \Gamma S_w^T \Omega = \Gamma_w \Omega$ will be equal to $\{ \pm i \sigma_i \}$ \cite{vidal02}. The $n$ absolute values $\{ \sigma_i \}$ of the elements of the spectrum correspond to the $n$ dinstinct eigenvalues of $\Gamma_w$. We say that these are the \textit{symplectic eigenvalues} of the (covariance) matrix $\Gamma$ and call the set $\{\sigma_i\}$ the \textit{symplectic spectrum}. This spectrum characterizes the Gaussian state. By construction it is invariant under any Gaussian transformation. \\
Once the covariance matrix $\Gamma$ of the state is given or calculated, the set of symplectic eigenvalues can be directly obtained from the spectrum of the matrix $M = \Gamma \Omega$. We will see in the next section how the symplectic eigenvalues contain the total information about entanglement properties of a  Gaussian state. This is the reason why this algebraic description of Gaussian states is efficient. It allows one to quantify entanglement simply from the symplectic eigenvalues of the matrix of correlations of the quadrature operators, which are in general much easier to calculate than the eigenvalues of the density matrix.


\section{Derivation of the formula}

In this section we give a step-by-step derivation of the formula (\ref{entropy}) for the von Neumann entropy of a Gaussian state, showing explicitly the connection between symplectic eigenvalues of the covariance matrix and entropy.\\
Consider a general Gaussian state $\rho = \sum_j p_j | \phi_j\9 \6 \phi_j|$ corresponding to $n$ modes with a covariance matrix $\Gamma$. We can always find a symplectic transformation $S$ such that $\rho^\prime = \hat{U}(S)\, \rho \, \hat{U}^{\dagger}(S)   \to \hr^\prime = S \hr$, which takes $\Gamma$ to the normal form $\Gamma^\prime = S \Gamma S^T$.

Note that in general this is not the Williamson form. Hence, we can rewrite $\rho^\prime$ as
\be
\label{decoupling}
\rho^\p = \rho_1^\prime \otimes \rho_2^\p ... \otimes \rho_n^\p \, ,
\ee
where each $\rho_i^\p$ is the density matrix of a single thermal oscillator \cite{botero03}. Observe that after the transformation, the transformed oscillators are now uncoupled non-local thermal oscillators. A 
harmonic oscillator in thermal equilibrium at temperature $T$ is described by a 
canonical ensemble \cite{cohen}. In the number basis $\{| \varphi_n \9\}$, its density matrix 
is written as \cite{kubo}
\be
\rho = \sum_{n} p_n  | \varphi_n \9 \6 \varphi_n | \, ,
\ee
where the $p_n$'s are the probabilities associated to each state and correspond to $p_n = Z^{-1} e^{- E_n /k_B T}$, with $E_n$ energies of the $n$-th state of the Hamiltonian $\hat{H} |\varphi_n \9 = E_n |\varphi_n \9$, and where the partition function $Z =  \text{Tr}\left( e^{- \hat{H} / k_B T} \right)$ is a normalization constant. Hence, the density matrix of each uncoupled mode is equal to
\bal
\rho_i^\prime &= \sum_n Z_i^{-1}e^{-E_{n,i}^\prime / k_B T_i} |\varphi_{\n} \9 \6 \varphi_{\n} | =\\
&= Z_i^{-1} e^{- \hat{H}_i^\p / k_B T_i} \,.
\end{align}
Using the transformed creation and annihilation operators
\bal
\label{ladder}
\ap = \sqrt{\frac{m_i^\p \omega_i^\p}{2 \hbar}} \left( \hat{q}^{\p}_i + \frac{i}{m_i^\p \omega_i^\p}\hat{p}_i^{\p}  \right)\,\,\,\,\,\,\,\,\,\,\,\,\,\,\, \adp = \sqrt{\frac{m_i^\p \omega_i^\p}{2 \hbar}} \left( \hat{q}^{\p}_i - \frac{i}{m_i^\p \omega_i^\p}\hat{p}_i^{\p}  \right)\, ,
\end{align}
whose transformation under $S$ follows from that for quadrature operators,
\be
 \xi = \left( \hat{a}_1, ..., \hat{a}_n, \hat{a}^\dagger_1, ..., \hat{a}^\dagger_n \right)^T \longrightarrow \xi^\prime = S \xi \, ,
\ee
we can rewrite the Hamiltonian of each oscillator as $\hat{H}_i^\p = \hbar \om (\adp \ap + \frac12)$ and the partition function $Z_i$ as
\be
Z_i = \text{Tr}\left( 
e^{- \hat{H}_i^\prime / k_B T_i} \right) = \sum_{\n=0}^{\infty} \6 \varphi_{\n} | e^{-\left( \adp \ap + \frac12 \right)\hbar \om/ k_B T_i} |\varphi_{\n} \9 = \sum_{\n=0}^\infty e^{- \left(n_i^\p +\frac12 \right) \beta_i} \, ,
\ee
where $\n$ is the eigenvalue of the number operator $\adp \ap | \varphi_{\n} \9 = \n |\varphi_{\n} \9$ and $\beta_i \equiv \hbar \om / k_B T_i$ is a cumulative parameter that depends on the transformed frequency $\om$. We can rewrite the last bit using the properties of the geometric series
\be
Z_i = e^{-\beta_i/2} \sum_{\n=0}^\infty e^{-n_i^\p \beta_i} = e^{- \beta_i/2} \left [ 1+ e^{-\beta_i} + e^{-2 \beta_i} + ... \right] \longrightarrow z_i= \frac{e^{-\beta_i/2}}{1 - e^{-\beta_i}} \, ,
\ee
and finally we have
\be
\label{newrho}
\rho_i^\p = \left(1 - e^{-\beta_i} \right) e^{-\adp \ap \beta_i} \, .
\ee
To simplify this expression we can rewrite the density matrix in terms of $\bn = \6 n_i^\p \9$, the mean occupation number of the transformed modes:
\bal
\label{n}
\notag
\bn &= \6 \adp \ap \9 = \text{Tr}(\rho_i^\p \adp \ap) = \sum_{\n=0}^\infty \6 \varphi_{\n} | \left(1 - e^{-\beta_i} \right) e^{-\adp \ap \beta_i} \adp \ap | \varphi_{\n} \9 = \\
&= \left(1 - e^{-\beta_i} \right) \sum_{\n=0}^\infty \n e^{- \n \beta_i} = \frac{1}{e^{\beta_i} -1} \longrightarrow  \, e^{\beta_i} = \frac{1 +\bn}{\bn} \, .
\end{align}
As a result we obtain the following expression for the density matrix
\bal
\label{lastrho}
\rho_i^\prime = \frac{1}{1 +\bn} \left(\frac{\bn}{1 +\bn} \right)^{\adp \ap} \, .
\end{align}
Calculation of the von Neumann entropy $S(\rho_i^\prime) = - \text{Tr}  \left( \rho_i^\prime \log{\rho_i^\prime} \right)$ gives:
\bal
\notag
\text{Tr} \left( \rho_i^\prime \log{\rho_i^\prime} \right) &= \sum_{\n =0}^\infty \6 \varphi_{\n} | \frac{1}{1+\bn}\left( \frac{\n}{1+\n} \right)^{\adp \ap} \log \left [ \frac{1}{1+\bn}\left( \frac{\n}{1+\n} \right)^{\adp \ap} \right] | \varphi_{\n} \9 =\\
\notag
& = \frac{1}{1+\bn} \sum_{\n =0}^\infty \6 \varphi_{\n} | \left( \frac{\bn}{1+\bn} \right)^{\n} \left[ - \log (1+\bn) + \n \log \left( \frac{\bn}{1+\bn} \right) \right] | \varphi_{\n} \9 = \\
\notag
& = \frac{1}{1+\bn} \left[ - \log (1+\bn) \sum_{\n =0}^\infty   \left( \frac{\bn}{1+\bn} \right)^{\n} + \log \left( \frac{\bn}{1+\bn} \right)  \sum_{\n =0}^\infty \n  \left( \frac{\bn}{1+\bn} \right)^{\n} \right] = \\
& = \bn \log \bn - (1+ \bn) \log (1+\bn) \, ,
\end{align}
where we used
\be
\sum_{\n =0}^\infty   \left( \frac{\bn}{1+\bn} \right)^{\n} = (1+\bn) \,\,\,\,\,\,\,\,\, \text{and} \,\,\,\,\,\,\,\,\, \sum_{\n =0}^\infty \n  \left( \frac{\bn}{1+\bn} \right)^{\n} = \bn (1+\bn) \, .
\ee
It immediately follows that for a single oscillator thermal state, the von Neumann entropy is expressed, in terms of mean occupation number, as
\be
\label{vne}
S(\rho_i^\prime) = (1+ \bn) \log (1+\bn) - \bn \log \bn \, ,
\ee
a well-known result in statistical physics \cite{kubo}.\\
The connection of the entropy with the symplectic eigenvalue $\sigma_i^\p$ of the state follows straightforwardly. After the symplectic transformation on the canonical variables that takes $\Gamma$ to its normal form, the covariance matrix of the reduced state $\rho_i^\prime$ looks like
\bal
\label{covi}
\Gamma_i^\prime = \left (
\begin{tabular}{c c}
$\6 \hq_i^{2 \, \prime} \9$ & $0$\\
$0$ & $\6 \hp_i^{2 \, \prime} \9$\\
\end{tabular} \right ) \, .
\end{align}
Using the ladder operators (\ref{ladder}) it is easy to show that
\bal
\6 \hq_i^{2 \, \prime} \9 = \text{Tr}(\rho_i^\prime \hq_i^{2 \, \prime} ) =& Z^{-1} \sum_{\n =0}^\infty \6 \varphi_{\n} | \frac{\hbar}{ 2 m_i^\prime \omega_i^\prime} (\adp+\ap)(\adp+\ap) e^{-\left( \adp \ap + \frac12 \right)\hbar \om/ k_B T_i} | \varphi_{\n} \9 =\\
=& \frac{\hbar}{2 m_i^\prime \omega_i^\prime} + \frac{\hbar}{m_i^\prime \omega_i^\prime} \frac{1}{e^{\beta_i} - 1} = \frac{\hbar}{2 m_i^\prime \omega_i^\prime} \coth{\frac{\beta_i}{2}} \, ,
\end{align}
and, proceeding in the same way, that
\be
\6 \hp_i^{2 \, \prime} \9 = \frac{\hbar m_i^\prime \omega_i^\prime}{2} + \hbar m_i^\prime \omega_i^\prime \frac{1}{e^{\beta_i} - 1} = \frac{\hbar m_i^\prime \omega_i^\prime}{2} \coth{\frac{\beta_i}{2}} \, .
\ee
Our previous discussion allows us to exhibit a relationship between the symplectic eigenvalue $\sigma_i^\prime$ of the system and the mean occupation number. We know that the spectrum of $\Gamma_i^\prime \Omega$ corresponds to $\{\pm i \sigma_i^\prime \}$, and therefore from (\ref{n}) we have that:
\bal
\text{eigenvalues}\, \{\Gamma_i^\prime \Omega\} = \pm i \sqrt{\6 \hq_i^{2 \, \prime} \9 \6 \hp_i^{2 \, \prime} \9} = \pm i \frac{\hbar}{2} \frac{e^{\beta_i} + 1}{e^{\beta_i} -1} = \pm i \hbar(\bn + \frac12) \,
\end{align}
and after fixing the units such that $\hbar = 1$, we find the equivalence
\be
\label{sn}
\sigma_i^\prime = \bn + \frac12 \longrightarrow \bn = \sigma_i^\prime -\frac12 \, .
\ee
Entropy is an additive quantity, therefore the total entropy of a state $\rho^\prime$ which is the direct tensor product of $n$ states is just the sum of the entropies of each state. It is easy to rewrite the formula (\ref{vne}) as a function of the symplectic eigenvalues,
\be
\label{entro}
S(\rho_i^\prime) = \sum_{i=0}^n \left [ \left( \sigma_i^\prime + \frac12 \right) \log_2 \left( \sigma_i^\prime + \frac12 \right)  - \left( \sigma_i^\prime - \frac12 \right) \log_2 \left( \sigma_i^\prime - \frac12 \right)  \right] \, .
\ee
Remind that the symplectic eigenvalues are the invariants of the correlation matrix. Since the entropy of a Gaussian state is solely a function of the symplectic eigenvalues, entropy itself is invariant under a symplectic transformation. This means that $S(\rho) = S(\rho^\prime)$. Hence we can drop the prime in the formula (\ref{entro}) and find (\ref{entropy}), which concludes the derivation.\\
We also notice from equations (\ref{n}) and (\ref{sn}) that the thermal parameter of each oscillator depends on the correspondent symplectic eigenvalue:
\be
\label{beta}
\beta_i = \ln \left( \frac{1 + \bn}{\bn} \right) =  \ln \left( \frac{\sigma_i + 1/2}{\sigma_i - 1/2} \right) \, .
\ee
\section{Conclusions}

We conclude
with an example and a couple of considerations about entanglement and its thermal properties.\\
Systems of coupled harmonic oscillators, like harmonic chains \cite{audenaert02, botero04}, are relatively well-understood and easy to describe mathematically. There exists an extensive literature about quantifying the bipartite entanglement of such systems. Among other reasons, this is the case because some of these entanglement measures may follow area laws \cite{entop}. States of coupled harmonic oscillators are also interesting because they can exhibit Gaussian properties. In particular, when the interaction between the modes is quadratic, the ground state of the system is Gaussian.

For a pure state the entanglement entropy and the von Neumann entropy coincide \cite{plenio06}. Thus we can use the formalism introduced earlier to describe the entanglement properties of a $n$ modes pure Gaussian state $\rho$ with covariance matrix $\Gamma$. We have seen that through a proper global symplectic transformation $S$, the state $\rho$ can be decomposed as the tensor product of $n$ single thermal oscillator (\ref{decoupling}). Now the new non-local states $\rho_i^\prime$ have to be pure as well for the properties of decomposition of a pure state. Hence the decoupled oscillator are all in their ground state and we assign to each of them a virtual temperature $T_i=0$. From equation (\ref{sn}) it follows that the symplectic eigenvalues are all equal to $\sigma_i = \frac{1}{2}$ and making use of (\ref{entropy}) we correctly find that the total entropy of any pure state $\rho$ is zero. \\
Divide now the $n$ modes into two sets $A = \{A_1, ..., A_a\}$ and $B = \{B_1, ..., B_b\}$ such that $a+b = n$ and hand them to Alice and Bob (this very last step is not strictly necessary, but it is always fun). In order to calculate the entanglement entropy $S(\rho_A) = S(\rho_B)$ we need to obtain the symplectic eigenvalues that belong to one of the two partitions. To start, assume that Alice wants to study her part of the system. In general, after the division, the reduced density matrix $\rho_A = \text{tr}_B \rho$ corresponds to a mixed state. Therefore a local decomposition by means of a local symplectic transformation $S_A$,
\be
\label{dec}
\rho_A^\prime = \rho^\prime_{1,T_1}... \otimes \rho^\prime_{r,T_r} \otimes \rho^\prime_{r+1,G_{r+1}} ... \otimes \rho^\prime_{a,G_a}
\ee
contains both thermal and ground state oscillators.\\
Alice's symplectic spectrum $\{ \sigma_{1,..,a}\}_A$ is obtained from the spectrum of $\Gamma_A \Omega_A$ where $\Gamma_A = \text{tr}_B \Gamma$ is the reduced covariance matrix of the set A. Then Alice can use her symplectic eigenvalues into the von Neumann entropy formula for Gaussian states (\ref{entropy}) and quantify the bipartite entanglement. Suppose that Alice obtained $s$ symplectic eigenvalues satisfying
\bal
\sigma_{1, ..., s} \ge \frac{1}{2}\, ,
\end{align}
and $a-s$ symplectic eigenvalues $\sigma_{s+1, ..., a} = \frac12$. This means that the local decomposition (\ref{dec}) of Alice's set of modes can be rewritten as
\bal
\rho_A^\prime = \rho^\prime_{1,T_1}... \otimes \rho^\prime_{s,T_s} \otimes \rho^\prime_{s+1,G_{s+1}} ... \otimes \rho^\prime_{a,G_a} \, .
\end{align}
Physically this corresponds to having $s$ transformed thermal oscillators with thermal parameter $\beta_i$ given by equation (\ref{beta}) and $a-s$ ground states oscillators. Notice that only the thermal oscillators contribute to the bipartite entanglement entropy.\\
Botero and Reznik made this construction rigorous. They proved in \cite{botero03} that, after identifying the two sets $A=\{ A_1, ..., A_a \}$ and $B=\{ B_1, ..., B_b\}$, it is always possible to write the Gaussian pure state $|\phi \9 \equiv | \phi\9_{A,B}$ as
\bal
| \phi \9_{A,B} = | \widetilde{\phi}_1 \9_{\tilde{A}_1 \tilde{B}_1} \otimes ... \otimes | \widetilde{\phi}_s \9_{\tilde{A}_s \tilde{B}_s} \otimes | 0 \9_{\tilde{A}_{s+1,...,a}} \otimes | 0 \9_{\tilde{B}_{s+1,...,b}} \, ,
\end{align}
where $\tilde{A} = \{ \tilde{A}_1, ..., \tilde{A}_a \}$ and $\tilde{B} = \{ \tilde{B}_1, ..., \tilde{B}_b \}$ are the transformed modes resulting from the application of local symplectic transformations on the set $A$ and $B$, and $s$ is equal to the number of symplectic eigenvalues associated to Alice's reduced covariance matrix $\Gamma_A$ or to Bob's one $\Gamma_B$. It means that the state $| \phi \9_{A,B}$ can be rewritten as the direct sum of $s$ two-mode squeezed states, where each mode belongs to a different partition of the system, and $n-2s$ oscillator ground states.\\ Each two-mode squeezed state $| \widetilde{\phi}_j \9_{\tilde{A}_j \tilde{B}_j}$ is given by the expression,
\be
| \widetilde{\phi}_j \9_{\tilde{A}_j \tilde{B}_j} = \frac{1}{\sqrt{Z_j}} \sum_n e^{- \beta_j n/2} | n \9_{\tilde{A}_j} | n \9_{\tilde{B}_j} \, ,
\ee
and the squeezing parameter $\beta_j$ corresponds to the thermal parameter of the $j$-th thermal oscillator of the local normal-modes decomposition of $\rho_A$ (or $\rho_B$). \\
We hope this dicussion will help to clarify the necessary steps to derive the entropy expression for Gaussian states and offer at the same time an extensive literature where the reader can find more details on the subject.

\subsection{Two coupled harmonic oscillators}
We want to present an easy case in order to show some explicit calculations. Imagine to have a system composed of two quantum harmonic oscillators with mass $m$ and frequency $\omega$ coupled in position and described by the following Hamiltonian:
\be
\hat{H} = \frac{1}{2m}(\hp_1^2 + \hp_2^2) + \frac{m \omega^2}{2} (\hq_1^2 + \hq_2^2) + \lambda (\hq_1 - \hq_2)^2 \, .
\ee
The ground state of this system is Gaussian. We want to calculate the bipartite entanglement between the two oscillators for the ground state. The global symplectic transformation $S$, described by the matrix
\bal
S =   \frac{1}{\sqrt{2}} \left (
\begin{tabular}{c c c c}
$1$ & $-1$ & $0$ & $0$\\
$1$ & $1$ & $0$ & $0$\\
$0$ & $0$ & $1$ & $-1$\\
$0$ & $0$ & $1$ & $1$\\
\end{tabular} \right ) \, ,
\end{align}
gives two uncoupled oscillators with the new frequencies
\be
\omega_1^\prime = \omega\,, \,\,\,\,\,\,\,\,\,\, \omega_2^\prime = \omega \sqrt{1 + \frac{4 \lambda}{m \omega^2}} \equiv \omega \alpha \,.
\ee
The corresponding covariance matrix for the normal modes, see (\ref{covi}),
\bal
\Gamma^\prime =   \frac{1}{\sqrt{2}} \left (
\begin{tabular}{c c c c}
$\frac{1}{2 m \omega}$ & $0$ & $0$ & $0$\\
$0$ & $\frac{1}{2 m \omega \alpha}$ & $0$ & $0$\\
$0$ & $0$ & $\frac{m \omega}{2}$ & $0$\\
$0$ & $0$ & $0$ & $\frac{m \omega \alpha}{2}$\\
\end{tabular} \right ) \, ,
\end{align}
can be transformed back using (\ref{gam}) into the covariance matrix $\Gamma$ of the system
\bal
\Gamma = S^{-1} \Gamma^\prime (S^T)^{-1} =   \frac{1}{\sqrt{2}} \left (
\begin{tabular}{c c c c}
$\frac{1+\alpha}{4 m \alpha \omega}$ & $\frac{1- \alpha}{4 m \alpha \omega}$ & $0$ & $0$\\
$\frac{1- \alpha}{4 m \alpha \omega}$ & $\frac{1+\alpha}{4 m \alpha \omega}$ & $0$ & $0$\\
$0$ & $0$ & $\frac{1}{4}m(1+\alpha)\omega$ & $\frac{1}{4}m(-1+\alpha)\omega$\\
$0$ & $0$ & $\frac{1}{4}m(-1+\alpha)\omega$ & $\frac{1}{4}m(1+\alpha)\omega$\\
\end{tabular} \right ) \, ,
\end{align}
It is straightforward to trace out the complementary degrees of freedom and obtain the reduced covariance matrix for the first (or equivalently, the second) oscillator
\bal
\Gamma_1 = \Gamma_2 = \left(
\begin{tabular}{c c}
$\frac{1+\alpha}{4 m \alpha \omega}$ & $0$\\
$0$ & $\frac{1}{4}m(1+\alpha)\omega$\\
\end{tabular} \right) \, .
\end{align}
The last step is to calculate the only symplectic eigenvalue of the reduced covariance matrix from the spectrum of $\Omega_{2 \times 2} \Gamma_{1,2}$, which is given by
\be
\sigma_1 = \sigma_2 = \frac{1 + \alpha}{4 \sqrt{\alpha}} \, ,
\ee
and finally it can be used to quantify the bipartite entanglement using the bipartite entanglement formula (\ref{entropy}).

\subsection{Acknowledgments}
The author is extremely grateful to Daniel Terno, Gavin Brennen and Trond (Thorn) Linjordet for reading the
manuscript at expenses of their own time and providing useful comments and criticisms.

\end{document}